\documentclass[12pt,sort&compress]{iopart}
\usepackage{upgreek}
\usepackage{xcolor}
\usepackage{graphicx}

\usepackage[numbers]{natbib}
\begin{document}
\title[Influence of source parameters on the phase-space distribution of a cryogenic beam]{Influence of source parameters on the longitudinal phase-space distribution of a pulsed cryogenic beam of barium fluoride molecules}

\author{M~C~Mooij$^{1,2}$, H~L~Bethlem$^{1,3}$, A~Boeschoten$^{2,3}$, A~Borschevsky$^{2,3}$, K~Esajas$^{2,3}$, T~H~Fikkers$^{2,3}$, S~Hoekstra$^{2,3}$, J~W~F~van Hofslot$^{2,3}$, K~Jungmann$^{2,3}$, V~R~Marshall$^{2,3}$, T~B~Meijknecht$^{2,3}$, R~G~E~Timmermans$^{2,3}$, A~Touwen$^{2,3}$, W~Ubachs$^{1}$, L~Willmann$^{2,3}$ and Y~Yin$^{2,3}$\footnote{Present address:
Department of Chemistry, University of Basel, Switzerland}\\ NL-\emph{e}EDM collaboration}

\address{$^1$ Department of Physics and Astronomy, LaserLaB, Vrije Universiteit Amsterdam, The Netherlands}
\address{$^2$ Nikhef, National Institute for Subatomic Physics, The Netherlands}
\address{$^3$ Van Swinderen Institute for Particle Physics and Gravity, University of Groningen, The Netherlands}
\ead{H.L.Bethlem@vu.nl}

\begin{abstract}
Recently, we have demonstrated a method to record the longitudinal phase-space distribution of a pulsed cryogenic buffer gas cooled beam of barium fluoride molecules. In this paper, we use this method to determine the influence of various source parameters. Besides the expected dependence on temperature and pressure, the forward velocity of the molecules is strongly correlated with the time they exit the cell, revealing the dynamics of the gas inside the cell. Three observations are particularly noteworthy: (1) The velocity of the barium fluoride molecules increases rapidly as a function of time, reaches a maximum 50-200\;$\upmu$s after the ablation pulse and then decreases exponentially. We attribute this to the buffer gas being heated up by the plume of hot atoms released from the target by the ablation pulse and subsequently being cooled down via conduction to the cell walls. (2) The time constant associated with the exponentially decreasing temperature increases when the source is used for a longer period of time, which we attribute to the formation of a layer of isolating dust on the walls of the cell. By thoroughly cleaning the cell, the time constant is reset to its initial value. (3) The velocity of the molecules at the trailing end of the molecular pulse depends on the length of the cell. For short cells, the velocity is significantly higher than expected from the sudden freeze model. We attribute this to the target remaining warm over the duration of the molecular pulse giving rise to a temperature gradient within the cell. Our observations will help to optimize the source parameters for producing the most intense molecular beam at the target velocity.
\end{abstract}

\noindent{\it Keywords\/}: Buffer gas cooled beam source, molecular beam, phase-space distribution, electric dipole moment of the electron

\maketitle

\section{Introduction}
Molecular radicals offer a number of unique possibilities for precision tests of fundamental physics theories~\cite{safronova2018a,andreev2018,roussy2023} and quantum technology~\cite{demille2002,ni2008,blackmore2019, yu2019}. Traditionally, these molecules are created using ovens~\cite{ramsey1985}, resulting in samples at relatively high temperatures which are of limited use for these applications. Rotationally and translationally cold samples of radicals have been produced by entraining laser-ablated species in a supersonic expansion of a carrier gas. In this way bright, $>10^{9}$ molecules per steradian per pulse in a single rotational level, and short, $<20\;\upmu$s, beam pulses have been generated, see for instance~\cite{aggarwal2021source} and references therein. The mean forward velocity of these beams is determined by the carrier gas and is typically around 600\;m/s when using argon and 300\;m/s for xenon~\cite{tarbutt2002}. 

A radically different approach for creating intense beams of molecules and molecular radicals is the so-called cryogenic buffer gas beam source, first introduced by Maxwell \emph{et al.}~\cite{maxwell2005} and further developed by Patterson \emph{et al.}~\cite{patterson2007}, van Buuren \emph{et al.}~\cite{vanbuuren2009}, Barry \emph{et al.}~\cite{barry2011}, Hutzler \emph{et al.}~\cite{hutzler2011, hutzler2012} and others. In this method, molecules are introduced into a cold cell by a capillary ~\cite{maxwell2005,patterson2007,patterson2009,vanbuuren2009,singh2018}, by laser ablation of a target containing a precursor~\cite{maxwell2005,patterson2007,hutzler2011,barry2011,lu2011,bulleid2013,hummon2013,zhou2015,santamaria2016,li2016,straatsma2017,bu2017,iwata2017,kozyryev2017b,albrecht2020,shaw2020} or by letting laser ablated atoms react with a donor gas~\cite{zhou2015,truppe2018,xu2019,esajas2021,hofsass2021}. The hot molecules are cooled by collisions with cold helium or neon buffer gas. After many collisions, a fraction of the molecules escapes the cell to form a molecular beam. The dimensions of the cell and the flow rate of the buffer gas determine the pressure within and are chosen in such a way that a significant fraction of the molecules is thermalized before hitting the wall of the cell. It was shown by Patterson \emph{et al.}~\cite{patterson2007} that by operating the source at a high flow rate, the molecules are more efficiently extracted from the cell, resulting in a more intense molecular beam. At the same time, the higher flow rate leads to a supersonic boost at the exit of the cell, resulting in a faster molecular beam. Typical beam intensities above 10$^{11}$ molecules per sr per pulse in a low rotational state have been reported with forward velocities below 170\;m/s~\cite{barry2011,hutzler2011}.
In order to have an efficient extraction while avoiding a supersonic boost to occur, two-stage cryogenic buffer cells have been investigated~\cite{patterson2007, lu2011}. Truppe \emph{et al.}~\cite{truppe2018} introduced a cell design that was optimized for creating relatively short ($\sim$300\;$\upmu$s) but intense molecular beams.

Over the last few years, we have constructed a pulsed cryogenic buffer gas cooled beam source that provides barium monofluoride (BaF) molecules for an experiment that will search for the electron's electric dipole moment (\emph{e}EDM)~\cite{aggarwal2018}. In the experiment, the BaF molecules will be decelerated to below 30\;m/s using a 4.5\;m long travelling-wave Stark decelerator~\cite{aggarwal2021deceleration}. There is a 
trade-off between the deceleration strength and the acceptance of the decelerator~\cite{vandemeerakker2012}. In order to decelerate a reasonable fraction of the beam with the current setup, we need a bright beam with an initial velocity not too far above 200\;m/s. Note that, the required length of the decelerator scales with the velocity squared; if the velocity of the initial beam would be 230\;m/s instead of 200\;m/s, we would need a decelerator with a length of 6\;m, to have the same acceptance. Therefore, understanding what determines the velocity of our source and how the source can be optimized for slow beams is of the utmost importance. 

Recently we have shown a sensitive method to determine the forward velocity of a buffergas cooled beam of barium fluoride molecules as a function of time after the ablation pulse, i.e. the longitudinal phase-space distribution of the beam, with high resolution~\cite{mooij2024}. Here we present a detailed study of the phase-space distribution as a function of various source parameters. Our paper is organized as follows: in section~\ref{sec:method}, we introduce our buffer gas beam source in detail, followed by an explanation of the so-called sudden freeze model that predicts the dependence of the velocity on the temperature and pressure of the buffer gas, and a recap of our detection method. 
In section~\ref{sec:characterization}, we present measurements of the phase-space distributions while varying source parameters such as the power of the ablation laser, the repetition rate, the temperature of the cell, the flow rate of the SF$_{6}$, the operation time, the flow rate of the neon buffer gas and the cell length. Finally, in section~\ref{sec:conclusions}, we summarize our findings.

\section{Method}\label{sec:method}

\subsection{Formation of the molecular beam}
\begin{figure}[t]
    \centering
    \includegraphics[width=0.66\columnwidth]{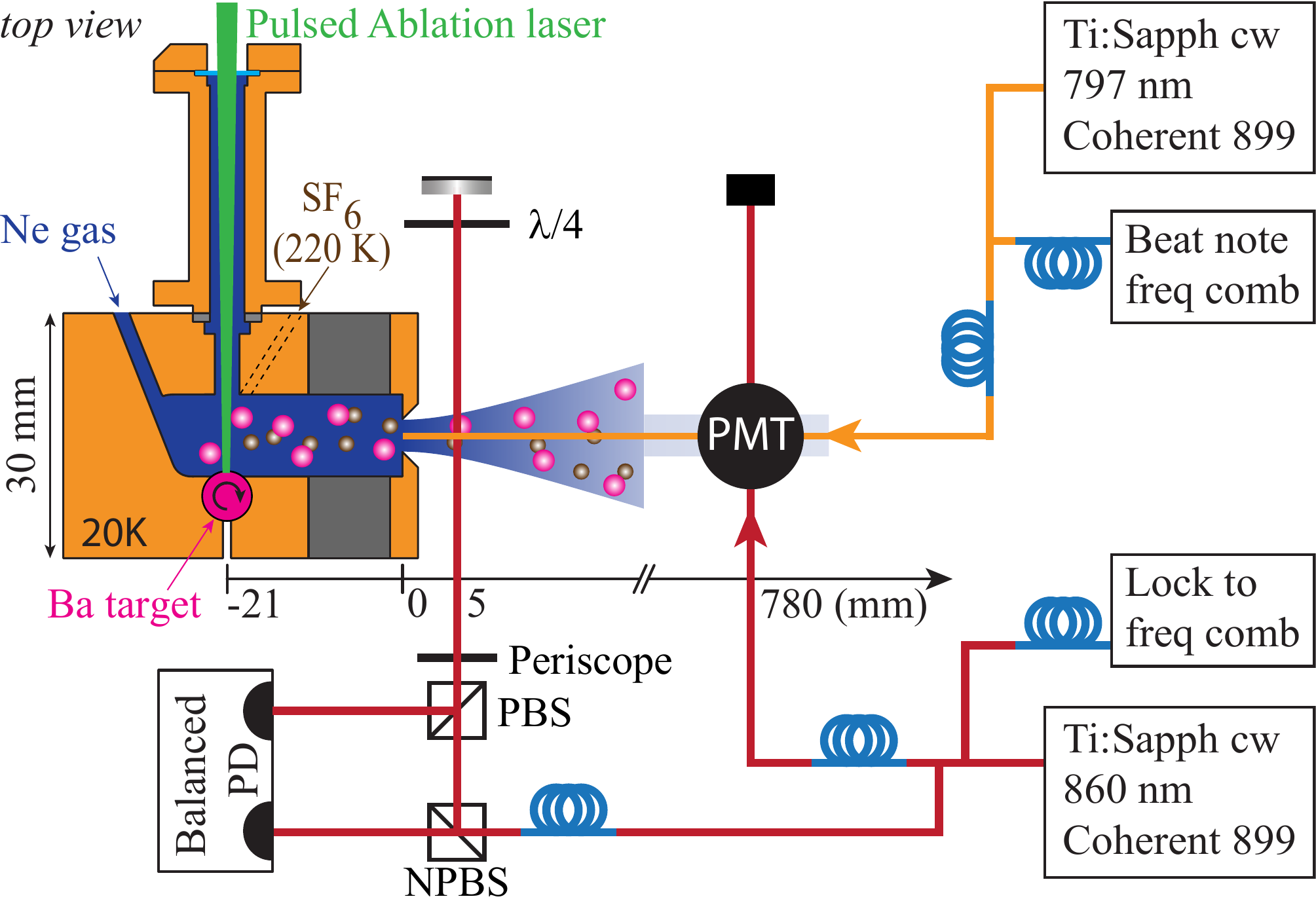}
    \caption{Schematic view of the experimental setup showing the cryogenic buffer gas beam source and the lasers used for absorption and fluorescence detection. Barium monofluoride molecules are created by letting sulphur hexafluoride molecules react with barium that is ablated from a solid barium rod inside a copper cell that is cooled to 20\;K. The molecules cool by collision with neon gas inside the cell, and expand through a 4.5\;mm diameter orifice to form a molecular beam. The distance between the target and the exit of the cell can be adjusted by inserting an extension tube between the cell body and the front plate. In the standard cell, the extension tube is absent and the distance between the target and the exit is 11\;mm, the figure depicts the situation when a 10\;mm long tube is inserted (indicated in gray). The phase-space distribution of the beam is recorded 780\;mm after the cell using a two-step laser excitation scheme. Fluorescence back to the ground state is measured using a photomultiplier tube (PMT).}
    \label{fig:setupVelocity}
\end{figure}
Figure~\ref{fig:setupVelocity} shows a schematic of our setup including a top-view of our cryogenic buffer gas beam source and the laser beam paths. The design of our cryogenic source is based on that of Truppe \emph{et al.}~\cite{truppe2018} and Esajas~\cite{esajas2021}. The heart of our setup is formed by a cubical copper cell kept at a temperature of around 20\;K using a 2-stage cryo-cooler (Sumitomo Heavy Industries, cold head RP-082B2S). A continuous flow of pre-cooled neon is passed through the cell with a flow rate of 10-70\;standard cubic centimeter per minute (\textsc{sccm})~\footnote{1\;\textsc{sccm} = 4.48$\times10^{17}$\;particles/s}.

The beam of BaF molecules is formed in four steps: 
(i) Inside the cell, barium is ablated by a pulsed Nd:YAG laser (532\;nm, 5\;ns pulse, 10\;Hz, typically 8\;mJ per pulse measured outside the vacuum) from a rotating solid Ba target. (ii) The barium atoms react with sulphur hexafluoride (SF$_{6}$) molecules that are injected into the cell from a copper tube kept at a temperature of 220\;K and with a flow rate of typically 0.03\;\textsc{sccm}. (iii) The BaF molecules created in this reaction are cooled via collisions with the neon atoms. (iv) They form a molecular beam by expanding through a 4.5\;mm diameter orifice into the vacuum. The cell is surrounded by a copper and an aluminium heat shield, at temperatures of 6\;K and $\sim$30\;K, respectively. These heat shields also provide the necessary pumping capacity to allow pressures on the order of 10$^{-2}$\;mbar inside the cell, while maintaining a pressure below 10$^{-6}$\;mbar in the molecular beam chamber.  

\subsection{Beam velocity as a function of pressure and temperature of the carrier gas}\label{subsec:theoryExpansion}

In this section, we will summarize some theory required to understand the dependence of the molecular beam velocity on the temperature and pressure of the buffer gas. If the source is operated at a very low buffer gas flow rate (for our cell well below 10\;\textsc{SCCM}) and hence low density, the effusive regime prevails and the molecules leave the exit without colliding with the buffer gas. Consequently, the velocity distribution of the molecules in the beam reflects the temperature inside the cell, corresponding to 52\;m/s for BaF molecules and 145\;m/s for neon atoms. 
In practice, we use a much higher flow rate to ensure that the molecules are entrained in the buffer gas and are pumped out of the cell before they have a chance to diffuse to the cell walls. In this so-called hydrodynamic regime many collisions between neon atoms and BaF molecules occur while they leave the cell. As the pressure inside the cell is higher than outside, these collisions lead to a net force that accelerates the neon atoms and barium fluoride molecules along the beam axis and lowers the longitudinal velocity spread of the beam. In this situation, the velocity of the molecules in the beam depends both on the temperature of the cell and the density of the neon gas. To make this statement more quantitative, we summarize here a derivation to relate the temperature, velocity and flow rate, which is based on the derivations of Pauly~\cite{pauly2000} and Hutzler \emph{et al.}~\cite{hutzler2011}. 

An isentropic expansion of a high-pressure gas into a vacuum leads to the conversion of internal energy into directed flow energy. When the distance from the source is large compared to the diameter of the exit of the cell, $z \gg d_{\mathrm{exit}}$, the density in the beam, $n(z)$ decreases quadratically with distance:

\begin{equation}
    n(z) \approx C \frac{n_0d_{\mathrm{exit}}^2}{z^2}
\end{equation} 
with $n_0$ the density in the cell and $C$ is a constant that is $\sim$0.25 for an effusive beam and $\sim$0.15 for a supersonic beam~\cite{pauly2000}. The density in the cell is a function of the flow rate, $f$, the size of the aperture $A=\pi d_{\mathrm{exit}}^2/4$, and the mean velocity of the beam, $\overline{v}$, and is given by~\cite{barry2011}:
\begin{equation}\label{eq:densityFlow}
    n_0 = \frac{4 f}{ A\overline{v}}.
\end{equation}

The decrease in density is accompanied by an increase in the most probable velocity,  $v_{\mathrm{mp}}(z)$, and a decrease in the temperature, $T(z)$, along the beam. For a mono-atomic gas such as neon, it can be derived that~\cite{depaul1993}:  

\begin{equation}
v_{\mathrm{mp}}(z) = v_{\infty} \left[1-\left(\frac{T(z)}{T_0}\right)\right]^{1/2} = v_{\infty} \left[1-\left(\frac{n(z)}{n_0}\right)^{2/3}\right]^{1/2}, 
\label{eq:isenExpT}
\end{equation}
with $T_0$ being the stagnation temperature of the gas in the cell. If all energy is converted, the temperature of the beam becomes zero, and the forward velocity becomes $v_{\infty}(T_0) = \sqrt{5k_{\mathrm{B}}T_0/m}$, with $m$ the mass of neon. In practice, the conversion process becomes increasingly slower while the density in the beam -- and hence the collision rate -- becomes smaller with distance from the exit of the cell. Consequently, the most probable velocity of the beam will be smaller than $v_{\infty}$. In the so-called `sudden freeze' model~\cite{pauly2000}, it is assumed that the expansion stops abruptly at a certain distance from the source, from which point the molecules travel in straight lines. The exact position, $z_0$, of this `quitting surface' is found by setting $Z_2$, the integral over the remaining two-body collisions after passing this surface, equal to 1:

\begin{equation}\label{eq:Z2}
    Z_2 = \int_{z_0/d_{\mathrm{exit}}}^{\infty} dZ_2 = 0.0465 \sqrt{\frac{8}{\pi}}\overline{\sigma}_{\mathrm{eff}} n_0 d^{8/3}_{\mathrm{exit}} z_0^{-5/3} = 1,
\end{equation}
with $\sigma_{\mathrm{eff}}$ being the effective (temperature averaged) cross-section. The numerical constant in the equation can be found from simulations~\cite{pauly2000}, but its exact value is unimportant for our purpose. Combining the above equations with~(\ref{eq:densityFlow}) that relates the density in the cell to the flow rate, $f$, we find:   

\begin{equation}\label{eq:fitVelFlow}
    v_{\mathrm{mp}}(f,T_0)  = v_{\infty}(T_0) \left[1-a\left( \frac{f }{v_{\infty}(T_0)} \right)^{-4/5}\right]^{1/2},
\end{equation}
with $a = 0.33 ( \overline{\sigma}_{\mathrm{eff}}/d_{\mathrm{exit}} )^{-4/5}$. 
This relation can be used to determine the temperature of the gas inside the cell, $T_0$, from measurements of the mean velocity as a function of the flow rate. \footnote{In this derivation, the flow velocity of the neon gas, which is 7-10 m/s in our experiments, is neglected.} 

\subsection{Detection}
To monitor the performance of the source, the BaF molecules are detected using absorption directly behind the cell on the $X^2\Sigma^+ \rightarrow A^2\Pi_{1/2}$ transition using $\sim 1\;\mu$W of light at 860\;nm in a 1\;mm diameter beam. The absorption signal can be converted into an absolute number by taking into account the spatial and velocity distributions of the beam in the longitudinal and transverse directions, using a procedure that is similar to the one described by Wright \emph{et al.}~\cite{wright2022}.
At the reference settings (to be discussed later), the peak absorption is typically 10\% (double pass), which corresponds to $1.9(6)\times10^{10}$\; BaF molecules in the $N=0$ state per pulse and $1.3(5)\times 10^{11}$\;molecules per sr per pulse. At a distance of 780\;mm from the source, the molecules are excited by light from two Ti:Sapphire lasers (Coherent 899) that are referenced to a frequency comb. One of the lasers is aligned perpendicular to the molecular beam and is resonant with the $X^2\Sigma^+ \rightarrow A^2\Pi_{1/2}$ transition at 860\;nm, while the other laser is aligned to be counter-propagating with respect to the molecular beam and resonant with the $A^2\Pi_{1/2} \rightarrow D^2\Sigma^+$ transition around 797\;nm. The frequency of this second laser is red-shifted with respect to the transition frequency to compensate for the Doppler shift. From this detuning, we infer the longitudinal velocity of the molecules. More information on this method can be found in~\cite{mooij2024}. 

Once excited to the $D$-state, part of the molecules will decay back to the ground state by emitting a photon at 413\;nm which is efficiently detected using a photomultiplier tube (PMT). A 40\;nm wide band-pass filter around 400\;nm is used to filter out scattered photons from the laser beams and unwanted fluorescence, resulting in a nearly background-free detection~\cite{murphree2009}.

\begin{figure}
    \centering
    \includegraphics[width=0.6\linewidth]{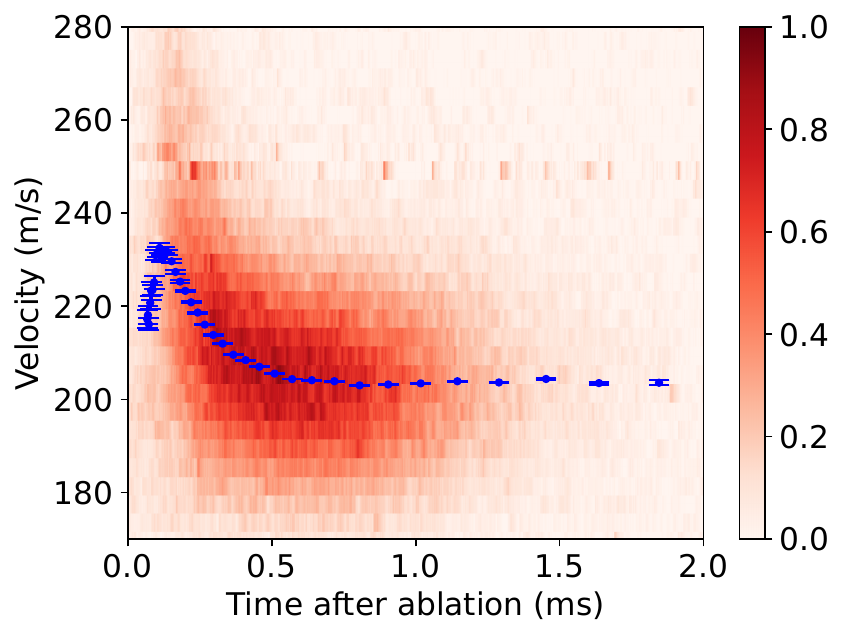}
    \caption{Phase space distribution of the molecular beam reconstructed at the exit of the source. The blue data points represent the mean velocities resulting from a Gaussian fit to the data at specific times. The horizontal line observed at 250\;m/s is due to difficulties in determining the frequency of the laser that drives the $A-D$ transition when its beat note with the frequency comb is equal to the repetition rate of the frequency comb.}
    \label{fig:meanVelocityFit}
\end{figure}

In order to be able to compare phase space distributions taken at different settings, we sum the reconstructed velocity distribution at the exit of the cell over a time interval that increases from 2\;$\upmu$s at the beginning of the pulse to about 250\;$\upmu$s at the end of the pulse and fit these with a Gaussian function. The mean velocity is shown as the blue data points that overlay the measured phase space distribution. The error bars represent the uncertainty from the fit.  

\section{Characterization of the cryogenic buffer gas beam source}\label{sec:characterization}

In this section, we will discuss the dependency of the longitudinal phase-space distribution of the molecular beam on various source parameters in order to understand the dynamics within the cell. 
We will vary these parameters one at a time around the values used for obtaining figure~\ref{fig:meanVelocityFit}, which we will refer to as the reference values. In section~\ref{subsec:ablation}, we will study the influence of the ablation power, the repetition rate, the temperature of the cell and the flow rate of the SF$_{6}$. In section~\ref{subsec:operationtime}, we will discuss the influence of the operation time. In section~\ref{subsec:flow}, we will discuss the influence of the neon flow rate and finally in section~\ref{subsec:celllength}, we will investigate how the phase space distribution of the beam changes when the length of the cell is increased.

\subsection{Influence of the ablation pulse energy and repetition rate, cell temperature and SF$_{6}$ flow rate}\label{subsec:ablation}

\begin{figure}
    \centering
    \includegraphics[width=0.9\linewidth]{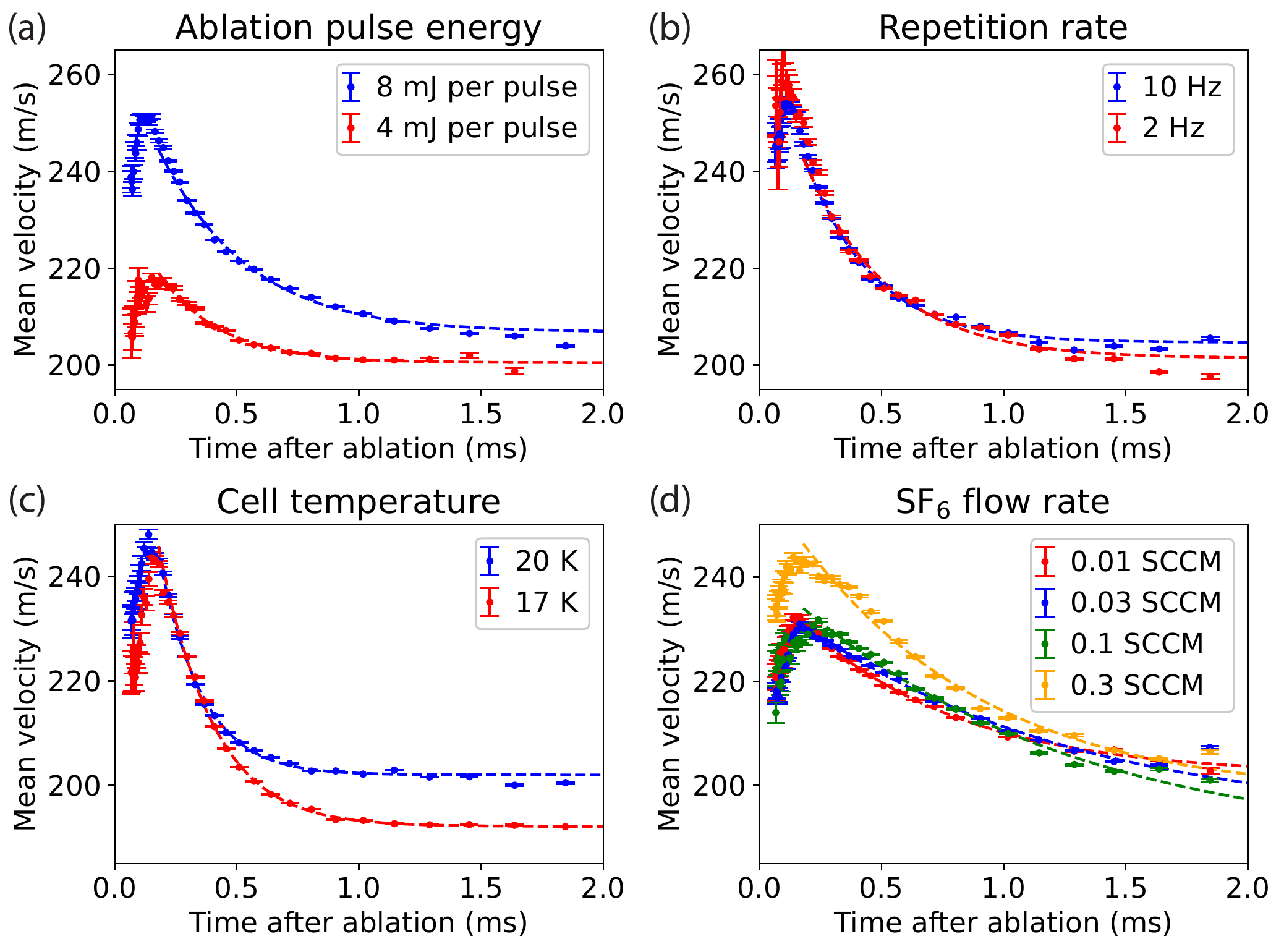}
    \caption{Mean velocity of the molecular beam as a function of time for different values of (a) the ablation pulse energy, (b) the repetition rate, (c) the cell temperature and (d) the SF$_{6}$ flow rate as indicated in the figure. All measurements are taken using a neon buffer gas flow rate of 20\;\textsc{sccm}. The blue data points in all sub-figures are different measurements taken at the reference settings. The dashed lines results from fits using~(\ref{eq:vExpDecay}).}
    \label{fig:ablationPulseEnergy}
\end{figure}

Figure~\ref{fig:ablationPulseEnergy}(a) shows the mean velocity of BaF molecules as a function of time after the ablation pulse while using ablation pulse energies of 8\;mJ (blue data points) and 4\;mJ (red data points). As observed, the velocity first increases and reaches a maximum $\sim$0.1\;ms after the ablation pulse and then decreases. At 8\;mJ, the velocity increases to above 250\;m/s, whereas at 4\;mJ, the velocity peaks at about 220\;m/s. The number of molecules exiting the source is decreased by about a factor of 3 when the pulse energy is decreased from 8\;mJ to 4\;mJ.

The fact that the velocity of the molecules is lower in the trailing end of the molecular beam does not seem surprising, given that barium fluoride molecules are produced at very high temperatures~\cite{davis1985} and require a minimum number of about 50\;collisions~\cite{decarvalho1999} before being cooled by the cold neon buffer gas. Naively, we may expect that the molecules that exit the cell shortly after the ablation pulse have had less collisions and are faster than molecules that exit the cell later. This is however \textit{NOT} what we see. Our measurements show that molecules leaving the cell immediately after the ablation pulse have comparable velocities to those at the tail of the pulse, while those that exit the cell 50-200\;$\upmu$s after the ablation pulse, are faster. We deduce from this that, \textit{at any time during the pulse}, the velocity of the BaF molecules reflects the temperature of the neon gas, i.e., the BaF molecules are always in thermal equilibrium with the buffergas. The observed correlation between the velocity and time reflects the sharp increase in temperature of the buffer gas due to the ablation plume and the subsequent decrease in temperature via conduction to the walls and collisions with the cold neon gas that is continuously flown into the cell. This hypothesis is consistent with the $\sim$50\;$\mu$s thermalisation time expected at the neon densities in our cell 
and also with earlier observations of Skoff \emph{et al.}~\cite{skoff2011}. Note that the cell body is not expected to heat up significantly by a single ablation pulse\footnote{From the mass of the cell body and the heat capacity of copper at 20~K~\cite{simon1992}, we estimate that the temperature of the cell body increases by $\sim$5\;mK due to a single laser pulse with an energy of 8\;mJ.}, 
hence we conclude that the limiting factor is the heat conduction of the neon gas to the cell walls. This will be discussed further in section~\ref{subsec:operationtime}. It may be observed that even after 1.5\;ms the mean velocity measured with an ablation energy of 8\;mJ per pulse is still slightly higher than that measured with 4\;mJ per pulse. We will come back to this difference in section~\ref{subsec:celllength}. 

Figure~\ref{fig:ablationPulseEnergy}(b) shows the mean velocity of BaF molecules as a function of time measured at the same ablation energy of 8\;mJ per pulse, but with a repetition rate of 10\;Hz (blue data points) or 2\;Hz (red data points). As expected, the measured velocities early in the pulse are very similar, but the velocity in the tail of the molecular pulse drops to a slightly lower velocity, indicating that the temperature of the cell may be slightly lower when operated at 2\;Hz instead of 10\;Hz. 

Figure~\ref{fig:ablationPulseEnergy}(c) shows the mean velocity of BaF molecules as a function of time measured when the copper cell is kept at 20\;K (blue data points) or 17\;K (red data points). Again, the measured velocities early in the pulse are very similar, but the velocity in the trailing end of the molecular pulse drops to a lower velocity when the cell is kept at a lower temperature. This shows that the source is ideally operated at the lowest possible cell temperature. The minimal temperature is determined by the requirement that the pressure anywhere in the system is above the vapour pressure of neon at that temperature. We observe that at a cell temperature below 20\;K, the neon line becomes completely clogged within $\sim$1\;hour, which we attribute to a small kink in the neon supply line. 

Figure~\ref{fig:ablationPulseEnergy}(d) shows the mean velocity of BaF molecules as a function of time measured when the SF$_{6}$ flow rate is varied from 0.01\;\textsc{sccm} to 0.3\;\textsc{sccm}. When the SF$_{6}$ flow rate is kept below 0.1\;\textsc{sccm}, the heat introduced by the SF$_{6}$ that is injected into the cell through a copper tube kept at a temperature of 220\;K, is apparently negligible. Note that the number of barium fluoride molecules produced with these flow rates is similar. 

If it is assumed that, after the initial rise, the temperature decreases exponentially, the velocity of the molecular pulse can be fitted to a simple exponential function of the form:  
\begin{equation} \label{eq:vExpDecay}
    v(t) = \sqrt{v_\mathrm{f}^2 + (v_\mathrm{i}^2 - v_\mathrm{f}^2)e^{-t/\tau} },
\end{equation}
with $v_\mathrm{i}$ and $v_\mathrm{f}$ being the initial and final velocities, respectively, and $\tau$ a characteristic time constant. These fits are shown as the dashed lines in figure~\ref{fig:ablationPulseEnergy}. As may be observed, the  
fitted characteristic time constants between the sets of measurements presented in panels (a)-(d) are rather different, which we blame on the cell being operated for extended times. In the next section, we will study this effect in more detail.

\subsection{Influence of the operation time}\label{subsec:operationtime}

\begin{figure}
    \centering
    \includegraphics[width=0.7\linewidth]{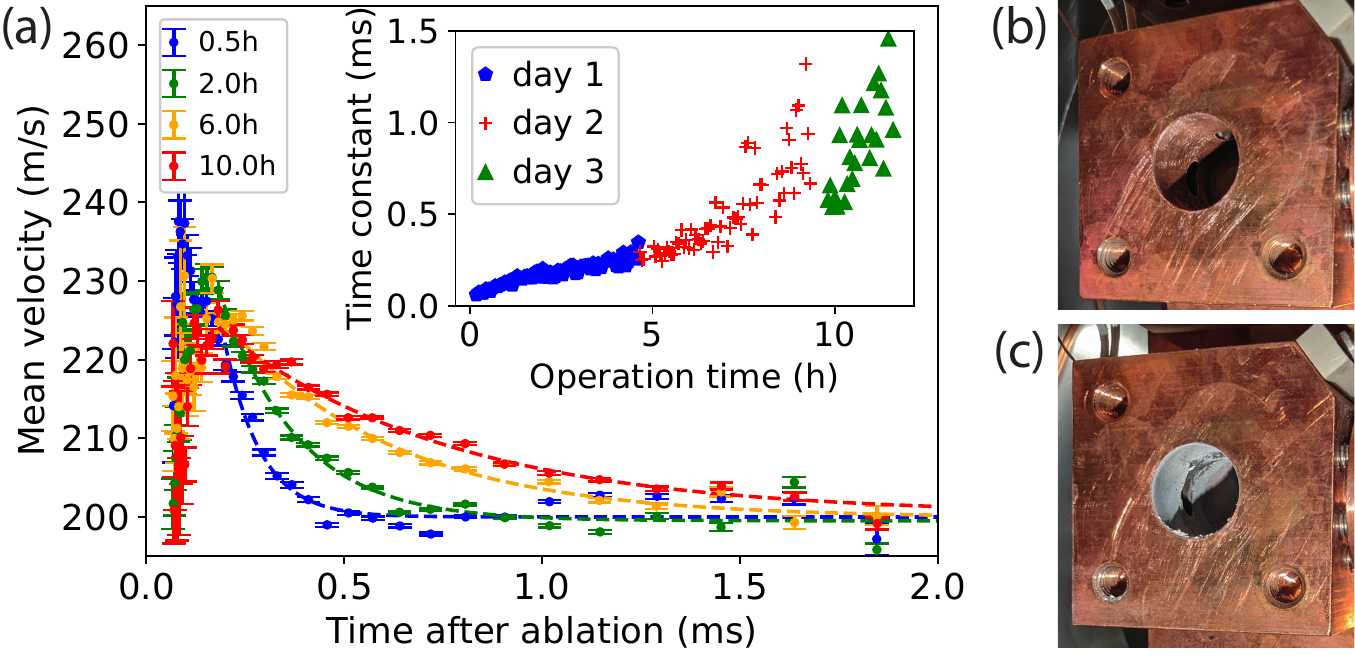}
    \caption{The effect of operation time. In (a), the mean velocity as a function of time is shown after operating the source for a time period as indicated. When the source is operated for extended times, the exponential decrease of the velocity becomes increasingly slower, resulting in a larger time constant as shown in the inset. (b) and~(c) are two pictures of the cell with the front plate removed that are taken before and after measuring the data shown in~(a), respectively. It is clear that the cell wall becomes covered with a substantial layer of dust.}
    \label{fig:time_constant}
\end{figure}

In order to study the effect of the operation time, we thoroughly cleaned the copper cell using acetic acid and subsequently measured the phase space distribution for up to 12 hours at the same setting.
After about 4.5\;hours of continuous operation and again after about 9\;hours, the source was heated up to 295\;K.
Figure~\ref{fig:time_constant}(a) shows measurements taken after operating the source for 0.5\;hour to 10\;hours. The dashed lines, also shown in the figure, are fits to the data using~(\ref{eq:vExpDecay}). The inset shows the resulting time constants from these fits as a function of operation time. During the approximately 12 hours of operation, the measured time constant increased from below 100\;$\upmu$s to above 1\;ms, which we attribute to barium, barium-sulfides and other reaction products, covering the walls of the cell. This dust decreases the thermal conductivity between the cell and the neon buffer gas. Cleaning the cell resets the source, while simply heating the cell to remove neon and SF$_{6}$ ice, has no effect on the measured velocity distribution. Figures~\ref{fig:time_constant}(b) and~(c) show photographs of the cell without front aperture, taken before and after measuring the date shown in figure~\ref{fig:time_constant}(a), respectively, clearly showing the contamination that builds up on the cell wall. The fact that the time constant changes during the operation of the source complicates the systematic study of the source considerably, forcing us to change parameters as quickly as possible while keeping an acceptable signal-to-noise ratio. Similar observations were made by Wright \emph{et al.}~\cite{wright2022} in their experiments on AlF. Using NF$_3$ instead of SF$_6$ as a fluor donor, they observed the same yield but with a significantly slower cell degradation. In our experiments on BaF, NF$_3$ gave a significantly smaller yield and was not studied further. 

\subsection{Influence of the neon flow rate}\label{subsec:flow}

\begin{figure}[t]
    \centering
    \includegraphics[width=\linewidth]{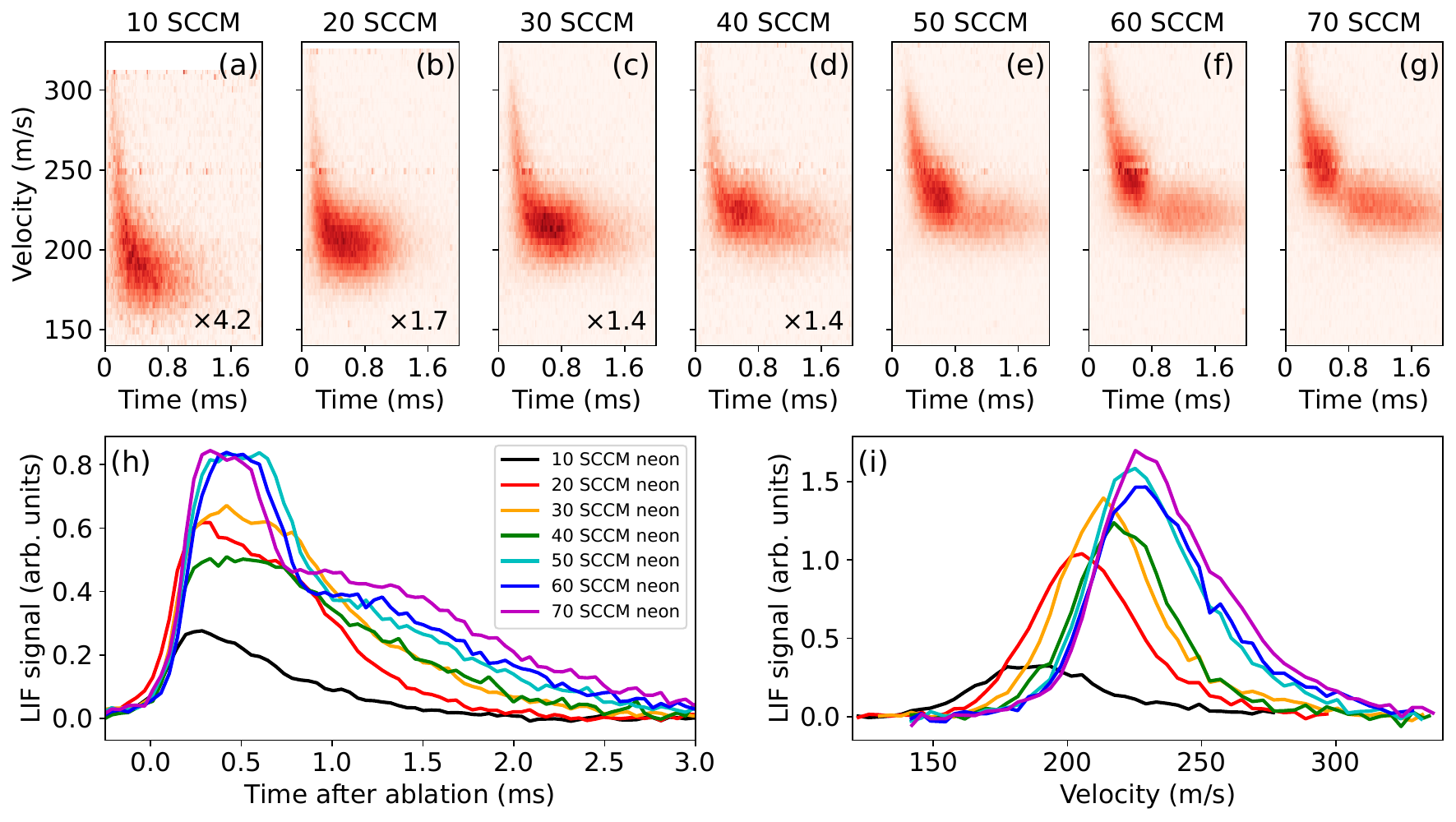}
    \caption{Phase-space distribution of the cryogenic beam for different neon flow rates. Panels~(a-g) show the reconstructed phase-space distribution at the source exit, where $t=0$ corresponds to the time that the ablation laser is fired. At low neon flow rates, the intensity has been multiplied with a factor as indicated in the panel. Panels (h) and (i) show the reconstructed intensity at different flow rates integrated over velocity and time, respectively. 
    A number of trends are observed with higher flow rate, as described in the main text.}
    \label{fig:expansion}
\end{figure}

In this section we will discuss the effect of the neon flow rate. Figure~\ref{fig:expansion} shows the reconstructed phase-space distributions at the exit of the cell for buffer gas flow rates between 10 and 70\;\textsc{sccm}~(a-g) along with the signal integrated over velocity~(h) or time~(i). 
Five trends are observed with higher flow rate: (i) The intensity and (ii) the pulse length of the molecular beam increase, as does the (iii) the characteristic time constant that describes the exponential decrease in temperature in the tail of the pulse. Furthermore, (iv) the mean velocity of the beam increases, while the (v) velocity spread decreases. The first two effects, which are most obvious from figure~\ref{fig:expansion}(h), are attributed to the increased diffusion time at higher neon densities, which reduces the loss of molecules frozen to the cell walls and leads to more efficient extraction from the cell. These effects have been discussed in detail by Patterson and Doyle~\cite{patterson2007}. The third effect is expected from the fact that the thermal diffusivity of the neon buffer gas decreases with increased density.
Finally, the fourth and fifth effects are due to the beam becoming more supersonic when the neon density in the cell is increased and will be analysed in more detail in the remainder of this section. Before doing this, there is one more observation worth noting. At higher flow rates, a dip in intensity is observed in the time of flight curves presented in figure~\ref{fig:expansion}(h) around 0.8\;ms after the ablation pulse. It has been suggested~\cite{TruppePrivateCommunication} that the molecules that exit the cell before (after) 0.8\;ms are formed from barium atoms that were ejected from the rod along (against) the direction of the buffer gas flow inside the cell. 





\begin{figure}
    \centering
    \includegraphics[width=\linewidth]{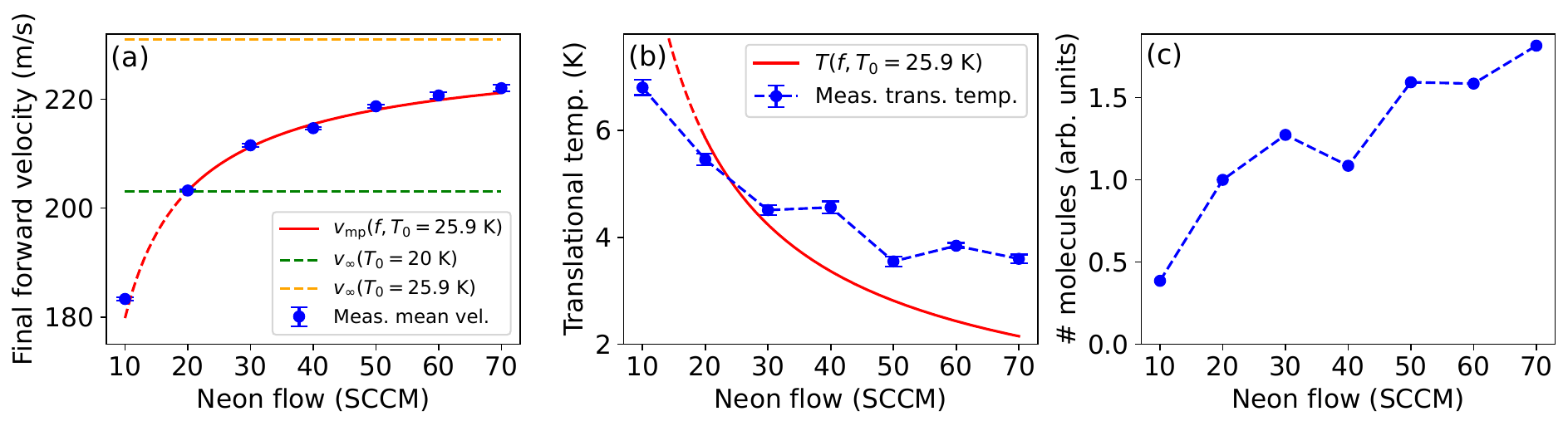}
    \caption{Velocity (a), temperature (b) and intensity (c) as a function of flow rate. (a) The blue data points show the final velocity, $v_\mathrm{f}$, in the tail of the molecular beam as a function of the neon flow rate. The solid/dashed red line shows a fit of~(\ref{eq:fitVelFlow}) to the data for flows between 20 and 70\;\textsc{sccm} from which we determine the terminal velocity $v_{\infty}$, shown as the dashed orange line, and the temperature of the buffer gas, $T_0$, which is 25.9\;K in this case. For completeness, the green dashed line shows the expected terminal velocity at a cell temperature of 20\;K. (b) The blue data points show the measured translational temperature, together with the prediction at 25.9\;K.
    (c) Number of BaF molecules in the $X^2\Sigma^+,v=0,N=0,J=1/2$ ground state.}
    \label{fig:fitTempFlow}
\end{figure}

As discussed in section~\ref{subsec:theoryExpansion}, the mean velocity and translational temperature of the molecular beam depend on the temperature and pressure of the neon buffer gas inside the cell, which in turn depends on the neon flow rate. The blue data points in figure~\ref{fig:fitTempFlow}(a) show the final forward velocity, $v_\mathrm{f}$, derived from fitting~(\ref{eq:vExpDecay}) to the mean velocity, as a function of time for each of the phase-space distributions presented in figures~\ref{fig:expansion}(a-g). The error bar of the data displays the uncertainty of this fit. The red solid line is a fit of~(\ref{eq:fitVelFlow}) to this data from which we determine $T_0$ and the terminal velocity $v_{\infty}(T_0)$, shown as the orange dotted line~\footnote{From this fit we find a collision cross-section $\sigma_{\mathrm{Ne-Ne}} = 1.9\times10^{-15}$\;cm$^2$, which is slightly larger than found by a simple hard-sphere model: $\sigma_{\mathrm{hs,Ne-Ne}} = 7.5\times10^{-16}$\;cm$^2$~\cite{pauly2000}.  }. In figure~\ref{fig:fitTempFlow}(b), the blue data points show the corresponding velocity spreads translated into a temperature. The solid red line shows the expected translational temperature from~(\ref{eq:isenExpT}) at $T_0$ found from the fit to the data in (a). As may be observed, the model fits the measured mean velocities (and to a lesser extent) the translational temperature well, however, from the fit parameter, $v_{\infty}(T)$, we find that the buffer gas temperature in the tail of the molecular beam is 25.9\;K, significantly above the temperature of the cell of 20\;K. This is a somewhat surprising result, given that on these time scales the buffer gas appears to have reached thermal equilibrium with the walls, while from the measurements performed at lower repetition rate, shown in figure~\ref{fig:ablationPulseEnergy}(b), it is seen that the temperature of the cell returned (close) to its set value within 100\;ms. We conclude from this that some part of the cell remains hot during the molecular pulse and only relaxes on a timescale 10-50\;ms. We believe that it is in fact the barium target that remains hot. More evidence for this will be presented in the next section. 


It would be insightful to determine the temperature of the buffer gas not only at the tail but at any time during the molecular beam pulse. However, this is complicated by the fact that the cooling rate towards the wall depends on the neon density in the cell and hence the flow rate. A rough estimate suggests that at its peak, the temperature of the buffer gas is increased to about 40\;K at an ablation pulse energy of 8\;mJ/pulse. 

Figure~\ref{fig:fitTempFlow}(c) shows the brightness of the beam of barium fluoride molecules in the $X^2\Sigma^+,v=0,N=0,J=1/2$ ground state as function of the neon flow rate found by integrating the phase-space distributions shown in figures~\ref{fig:expansion}a-g. As may be observed, the number of molecules increases by about a factor of 4.7 when the neon flow rate is increased from 10 to 70\;\textsc{sccm}. Note that, the sudden freeze model (presented in section~\ref{subsec:theoryExpansion}) predicts that the intensity does not depend strongly on the neon flow rate, as the slight increase of the relative population in the $N=0$ state due to the lower rotational temperature at high neon flow rates, is compensated by a the increased divergence of the beam~\cite{barry2011}. The observed intensity increase at higher flow rate is attributed to the increased efficiency in extraction of the molecules from the cell~\cite{patterson2007}. The fact that the forward velocity and intensity follow the same trend is a coincident. 

\begin{figure}[t]
    \centering
    \includegraphics[width=\linewidth]{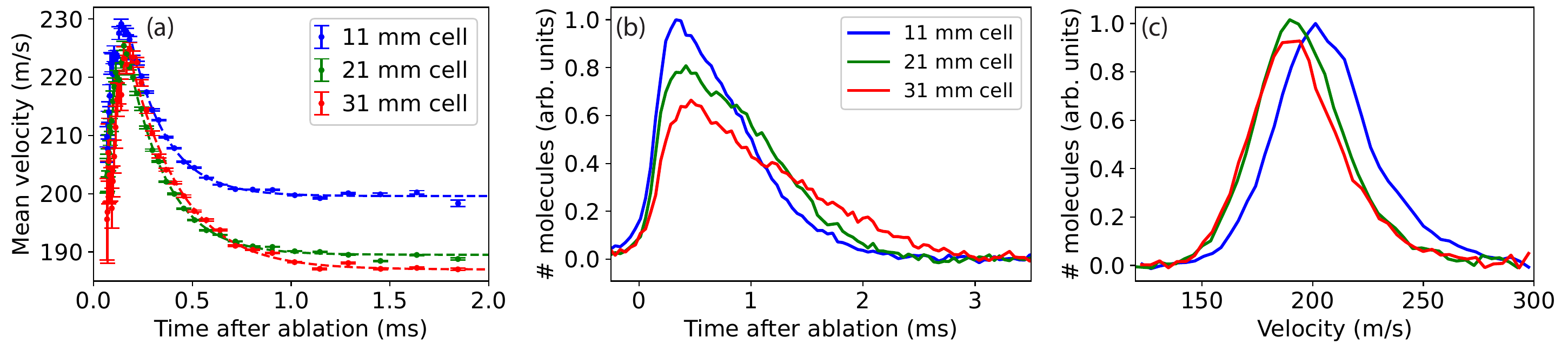}
    \caption{(a) Mean velocity as a function of time, (b) time-of-flight and (c) velocity distribution for three different cell lengths. The velocity in the tail of the molecular pulse is seen to decrease significantly, while the intensity is comparable.}
    \label{fig:extension}
\end{figure}

\subsection{Influence of the cell length}\label{subsec:celllength}
So far, all measurements presented in the paper were performed with the standard cell that has a distance of 11\;mm between the ablation target and the source exit. In this section, we will study what happens when the cell is extended to 21 or 31\;mm. Figure~\ref{fig:extension}(a) presents the mean velocity as a function of time, while (b) and~(c) present the intensity integrated over velocity or time, respectively. The ablation power, cell temperature, neon and SF$_{6}$ flow rates are set to the reference values. All measurement were taken after the cell was operated for 2 hours. Increasing the cell length to 21\;mm results in a longer molecular pulse that has a significantly lower velocity at the tail of the molecular pulse, while the number of molecules in the beam is comparable. Extending the cell further reduces the velocity but leads to a drop in the number of molecules by about 10\;\%. 
Note that, with a neon flow rate of 20\;\textsc{sccm}, the expected mean velocity for molecules exiting our cell at 20\;K is 178\;m/s. 

We attribute the observed dependence of the velocity on the length of the cell on the occurrence of a heat gradient in the cell. After the ablation pulse, the heat deposited in the ablated barium atoms is transferred to the buffer gas within 100\;$\upmu$s and is subsequently cooled away by the cell within a few ms. On the other hand, the heat deposited in the barium rod is transferred to the buffer gas at a slower rate and the target remains at elevated temperatures during the entire molecular pulse. Increasing the distance between the target and the exit results in the neon gas at the exit being closer to the cell temperature and the molecular beam being slower. This effect also explains the observed dependence of the final velocity on the ablation power, as was discussed in section~\ref{subsec:ablation}.  

\section{Conclusions}\label{sec:conclusions}

We presented measurements of the phase-space distribution of a cryogenic buffer gas beam of BaF. We observe a strong correlation of the mean forward velocity of the BaF molecules at the time they exit the source which is attributed to the neon buffer gas being warmed up by the plume of hot atoms released from the target by the ablation pulse and subsequently being cooled down via conduction to the cell walls. When the cell is operated for a longer period of time, the walls of the cell become covered with a layer of isolating dust which increases the time constant associated with the exponentially decreasing temperature of the neon gas. The barium target remains at elevated temperatures on a much longer time scale, resulting in a higher mean velocity than expected from the sudden freeze model. This velocity can be lowered by extending the length of the cell. Some of the observations above have been reported before, but our new method to accurately measure the phase-space distribution of the beam in combination with the stability and reproducibility of our cell allowed us to analyse these effects in great detail. As optimization of the source amounts to a compromize between brightness and a low forward velocity, a good understanding of the heating processes is pivotal for an optimal choice of the source parameters. In future work, we plan to also study the effect of the size and shape of the cell exit.  

\ack{The NL-\emph{e}EDM consortium receives program funding (EEDM-166 and XL21.074) from the Netherlands Organisation for
Scientific Research (NWO). We thank Johan Kos, Rob Kortekaas and Leo Huisman for technical assistance to the experiment. We acknowledge fruitful discussions with Mike Tarbutt and Stefan Truppe in the design of the cryogenic source.}

\bibliography{references}
\end{document}